\documentclass{article}
\usepackage[accepted]{icml2020}
\usepackage{microtype}
\usepackage{graphicx}
\usepackage{booktabs}

\usepackage{hyperref} 
\usepackage{amsmath}
\usepackage{pgfplots}
\usepackage{siunitx} 
\usepackage{subfig}
\pgfplotsset{compat=1.14}
\icmltitlerunning{Flowtron: an Autoregressive Flow-based Generative Network for Text-to-Speech Synthesis}
\begin{document}
\twocolumn[
\icmltitle{Flowtron: an Autoregressive Flow-based Generative Network for Text-to-Speech Synthesis}

\begin{icmlauthorlist}
\icmlauthor{Rafael Valle}{NVIDIA}
\icmlauthor{Kevin Shih}{NVIDIA}
\icmlauthor{Ryan Prenger}{NVIDIA}
\icmlauthor{Bryan Catanzaro}{NVIDIA}
\end{icmlauthorlist}

\icmlaffiliation{NVIDIA}{NVIDIA Applied Deep Learning Research (ADLR)}
\icmlcorrespondingauthor{Rafael Valle}{rafaelvalle@nvidia.com}

\icmlkeywords{Deep Learning, Generative Models, Text-to-Speech Synthesis}
\vskip 0.3in
]

\printAffiliationsAndNotice{}

\begin{abstract}
In this paper we propose Flowtron: an autoregressive flow-based generative network for text-to-speech synthesis with control over speech variation and style transfer. Flowtron borrows insights from IAF and revamps Tacotron in order to provide high-quality and expressive mel-spectrogram synthesis. Flowtron is optimized by maximizing the likelihood of the training data, which makes training simple and stable. Flowtron learns an invertible mapping of data to a latent space that can be manipulated to control many aspects of speech synthesis (pitch, tone, speech rate, cadence, accent). Our mean opinion scores (MOS) show that Flowtron matches state-of-the-art TTS models in terms of speech quality. In addition, we provide results on control of speech variation, interpolation between samples and style transfer between speakers seen and unseen during training. Code and pre-trained models will be made publicly available at \href{https://github.com/NVIDIA/flowtron}{https://github.com/NVIDIA/flowtron}.
\end{abstract}

\icmlkeywords{Audio Synthesis, Text-to-speech, Generative models, Deep Learning}

\section{Introduction} \label{sec:introduction}
Current speech synthesis methods do not give the user enough control over how speech actually sounds.
Automatically converting text to audio that successfully communicates the text was achieved a long time ago \cite{umeda1968synthesis, badham1983wargames}.
However, communicating only the text information leaves out all of the acoustic properties of the voice that convey much of the meaning and human expressiveness.
Nearly all the research into speech synthesis since the 1960s has focused on adding that non-textual information to synthesized speech.
But in spite of this, the typical speech synthesis problem is formulated as a text to speech problem in which the user inputs only text.

Taming the non-textual information in speech is difficult because the non-textual is unlabeled. 
A voice actor may speak the same text with different emphasis or emotion based on context, but it is unclear how to label a particular reading.
Without labels for the non-textual information, models have fallen back to unsupervised learning.
Recent models have achieved nearly human-level quality, despite treating the non-textual information as a black box. 
The model's only goal is to match the patterns in the training data \cite{shen2017natural,arik2017deep,arik2017deep2,ping2017deep}. Despite these models' excellent ability to recreate the non-textual information in the training set, the user has no insight into or control over the non-textual information. 

It is possible to formulate an unsupervised learning problem in such a way that the user can gain insights into the structure of a data set.
One way is to formulate the problem such that the data is assumed to have a representation in some latent space, and have the model learn that representation.
This latent space can then be investigated and manipulated to give the user more control over the generative model's output.
Such approaches have been popular in image generation for some time now, allowing users to interpolate smoothly between images and to identify portions of the latent space that correlate with various features \cite{radford2015unsupervised, kingma2018glow}.  

In audio, however, approaches have focused on embeddings that remove a large amount of information and are obtained from assumptions about what is interesting.  Recent approaches that utilize deep learning for expressive speech synthesis combine text and a learned latent embedding for prosody or global style \cite{wang2018style, skerry2018towards}. A variation of this approach is proposed by \cite{hsu2018hierarchical}, wherein a Gaussian mixture model (GMM) encoding the audio is added to Tacotron to learn a latent embedding. These approaches control the non-textual information by learning a bank of embeddings or by providing the target output as an input to the model and compressing it. 
However, these approaches require making assumptions about the dimensionality of the embeddings before hand and are not guaranteed to contain all the non-textual information it takes to reconstruct speech, including the risk of having dummy dimensions or not enough capacity, as the appendix sections in \cite{wang2018style,skerry2018towards, hsu2018hierarchical} confirm. 
They also require finding an encoder and embedding that prevents the model from simply learning a complex identity function that ignores other inputs. Furthermore, these approaches focus on fixed-length embeddings under the assumption that variable-length embeddings are not robust to text and speaker perturbations. Finally, most of these approaches do not give the user control over the degree of variability in the synthesized speech.

In this paper we propose Flowtron: an autoregressive flow-based generative network for mel-spectrogram synthesis with control over acoustics and speech. Flowtron learns an invertible function that maps a distribution over mel-spectrograms to a latent $\boldsymbol{z}$ space parameterized by a spherical Gaussian. With this formalization, we can generate samples containing specific speech charateristics manifested in mel-space by finding and sampling the corresponding region in $\boldsymbol{z}$-space. In the basic approach, we generate samples by sampling a zero mean spherical Gaussian prior and control the amount of variation by adjusting its variance. Despite its simplicity, this approach offers more speech variation and control than Tacotron.

In Flowtron, we can access specific regions of mel-spectrogram space by sampling a posterior distribution conditioned on prior evidence from existing samples \cite{kingma2018glow, gambardella2019transflow}. This approach allows us to make a monotonous speaker more expressive by computing the region in $z$-space associated with expressive speech as it is manifested in the prior evidence. Finally, our formulation also allows us to impose a structure to the $\boldsymbol{z}$-space and parametrize it with a Gaussian mixture, for example. In this approach related to \cite{hsu2018hierarchical}, speech charateristics in mel-spectrogram space can be associated with individual components. Hence, it is possible to generate samples with specific speech characteristics by selecting a component or a mixture thereof \footnote{What is relevant statistically might not be perceptually.}.

Although VAEs and GANs \cite{hsu2018hierarchical,binkowski2019high, akuzawa2018expressive} based models also provide a latent prior that can be easily manipulated, in Flowtron this comes at no cost in speech quality nor optimization challenges.

We find that Flowtron is able to generalize and produce sharp mel-spectrograms by simply maximizing the likelihood of the data while not requiring any additional Prenet or Postnet layer \cite{wang2017tacotron}, nor compound loss functions required by most state of the art models like \cite{shen2017natural,arik2017deep,arik2017deep2,ping2017deep,skerry2018towards, wang2018style,binkowski2019high}.

Flowtron is optimized by maximizing the likelihood of the training data, which makes training simple and stable. It learns an invertible mapping of the a latent space that can be manipulated to control many aspects of speech synthesis. Our mean opinion scores (MOS) show that Flowtron matches state-of-the-art TTS models in terms of speech quality. In addition, we provide results on control of speech variation, interpolation between samples, and style transfer between seen and unseen speakers with similar and different sentences. To our knowledge, this work is the first to show evidence that normalizing flow models can also be used for text-to-speech synthesis. We hope this will further stimulate developments in normalizing flows.
\section{Related Work}\label{sec:related_work}
Earlier approaches to text-to-speech synthesis that achieve human like results focus on synthesizing acoustic features from text, treating the non-textual information as a black box. \cite{shen2017natural,arik2017deep,arik2017deep2,ping2017deep}. Approaches like \cite{wang2017tacotron,shen2017natural} require adding a critical Prenet layer to help with convergence and improve generalization \cite{wang2017tacotron}. Furthermore, such models require an additional Postnet residual layer and modified loss to produce "\textit{better resolved harmonics and high frequency formant structures, which reduces synthesis artifacts}."

One approach to dealing with this lack of labels for underlying non-textual information is to look for hand engineered statistics based on the audio that we believe are correlated with this underlying information.

This is the approach taken by models like \cite{Nishimura2016SVS,lee2019adversarially}, wherein utterances are conditioned on audio statistics that can be calculated directly from the training data such as $F_0$ (fundamental frequency). However, in order to use such models, the statistics we hope to approximate must be decided upon a-priori, and the target value of these statistics must be determined before synthesis.

Another approach to dealing with the issue of unlabeled non-textual information is to learn a latent embedding for prosody or global style. This is the approach taken by models like \cite{skerry2018towards,wang2018style}, wherein in a bank of embeddings or a latent embedding space of prosody is learned from unlabelled data. While these approaches have shown promise, manipulating such latent variables only offers a coarse control over expressive characteristics of speech. 

A mixed approach consists of combining engineered statistics with latent embeddings learned in an unsupervised fashion. This is the approach taken by models like Mellotron \cite{valle2019mellotron}. In Mellotron, utterances are conditioned on both audio statistics and a latent embedding of acoustic features derived from a reference acoustic representation. Despite its advantages, this approach still requires determining these statistics before synthesis.
\section{Flowtron}\label{sec:audio_flow}
Flowtron is an autoregressive generative model that generates a sequence of mel spectrogram frames $p(x)$ by producing each mel-spectrogram frame based on previous mel-spectrogram frames $p(x) = \prod p(x_t \vert x_{1:t-1})$. Our setup uses a neural network as a generative model by sampling from a simple distribution $p(\boldsymbol{z})$. We consider two simple distributions with the same number of dimensions as our desired mel-spectrogram: a zero-mean spherical Gaussian and a mixture of spherical Gaussians with fixed or learnable parameters.

\begin{gather}
\boldsymbol{z} \sim \mathcal{N}(\boldsymbol{z};0,\boldsymbol{I}) \\
\boldsymbol{z} \sim \sum_k \hat{\phi}_k\, \mathcal{N}(\boldsymbol{z};\hat{\boldsymbol{\mu}_k},\hat{\boldsymbol{\Sigma}_k})
\end{gather}

These samples are put through a series of invertible, parametrized transformations $\boldsymbol{f}$, in our case affine transformations that transform $p(\boldsymbol{z})$ into $p(x)$.
\begin{gather}
\boldsymbol{x} = \boldsymbol{f}_0 \circ \boldsymbol{f}_1 \circ \ldots \boldsymbol{f}_k(\boldsymbol{z})
\end{gather}

As it is illustrated in \cite{kingma2016improved}, in autoregressive normalizing flows the $t$-th variable $\boldsymbol{z}^\prime_{t}$ only depends on previous timesteps $\boldsymbol{z}_{1:t-1}$:
\begin{gather}
\boldsymbol{z}^\prime_{t} = \boldsymbol{f}_{k} (\boldsymbol{z}_{1:t-1})
\end{gather}

By using parametrized affine transformations for $\boldsymbol{f}$ and due to the autoregressive structure, the Jacobian determinant of each of the transformations $\boldsymbol{f}$ is lower triangular, hence easy to compute. With this setup we can train Flowtron by maximizing the log-likelihood of the data, which can be done using the change of variables:
\begin{gather}
\log{p_\theta(\boldsymbol{x})} = \log{p_\theta(\boldsymbol{z})} + \sum_{i=1}^{k} \log
|\det(\boldsymbol{J}(\boldsymbol{f}_i^{-1}(\boldsymbol{x})))| \\
\boldsymbol{z} = \boldsymbol{f}_k^{-1} \circ \boldsymbol{f}_{k-1}^{-1} \circ \ldots \boldsymbol{f}_0^{-1}(\boldsymbol{x})
\end{gather}

For the forward pass through the network, we take the mel-spectrograms as vectors and process them through several ``steps of flow” conditioned on the text and speaker ids. A step of flow here consists of an affine coupling layer, described below.

\subsection{Affine Coupling Layer}
Invertible neural networks are typically constructed using coupling layers~\cite{dinh2014nice, dinh2016density, kingma2018glow}.  In our case, we use an affine coupling layer~\cite{dinh2016density}. Every input $\boldsymbol{x}_{t-1}$ produces scale and bias terms, $\boldsymbol{s}$ and $\boldsymbol{b}$ respectively, that affine-transform the succeeding input $\boldsymbol{x}_{t}$:  

\begin{gather}
(\log \boldsymbol{s}_t, \boldsymbol{b}_t) = NN(\boldsymbol{x}_{1:t-1}, \boldsymbol{text}, \boldsymbol{speaker}) \\
\boldsymbol{x}^\prime_t = \boldsymbol{s}_t \odot \boldsymbol{x}_t + \boldsymbol{b}_t
\end{gather}

Here $NN()$ can be any autoregressive causal transformation. This can be achieved by time-wise concatenation of a 0-valued vector to the input provided to $NN()$. The affine coupling layer preserves invertibility for the overall network, even though $NN()$ does not need to be invertible. This follows because the first input of $NN()$ is a constant and due to the autoregressive nature of the model the scaling and translation terms $\boldsymbol{s}_t$ and $\boldsymbol{b}_t$ only depend on $\boldsymbol{x}_{1:t-1}$ and the fixed $text$ and $speaker$ vectors. Accordingly, when inverting the network, we can compute $\boldsymbol{s_t}$ and $\boldsymbol{b_t}$ from the preceding input $\boldsymbol{x}_{1:t-1}$, and then invert $\boldsymbol{x}^\prime_t$ to compute $\boldsymbol{x}_t$, by simply recomputing $NN(\boldsymbol{x}_{1:t-1}, \boldsymbol{text}, \boldsymbol{speaker})$. 

With an affine coupling layer, only the $\boldsymbol{s_t}$ term changes the volume of the mapping and adds a change of variables term to the loss. This term also serves to penalize the model for non-invertible affine mappings.
\begin{equation}
\log |\det(\boldsymbol{J}(\boldsymbol{f}_{coupling}^{-1}(\boldsymbol{x})))| = \log |\boldsymbol{s}|
\end{equation}

With this setup, it is also possible to revert the ordering of the input $\boldsymbol{x}$ without loss of generality. Hence, we choose to revert the order of the input at every even step of flow and to maintain the original order on odd steps of flow. This allows the model to learn dependencies both forward and backwards in time while remaining causal and invertible.

\subsection{Model architecture}
\begin{figure}[!ht]
    \centering
    \includegraphics[width=\linewidth]{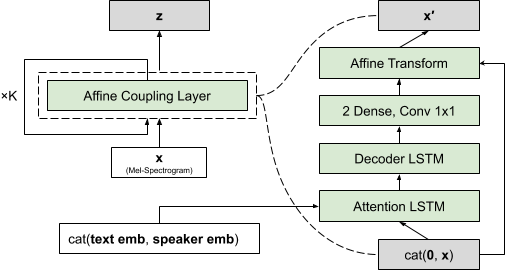}
    \caption{Flowtron network. Text and speaker embeddings are channel-wise concatenated. A 0-valued vector is concatenated with $\boldsymbol{x}$ in the time dimension.}
    \label{fig:Flowtron}
\end{figure}

Our text encoder modifies Tacotron's by replacing batch-norm with instance-norm. Our decoder and $NN$ architecture, depicted in Figure \ref{fig:Flowtron}, removes the \textit{essential} Prenet and Postnet layers from Tacotron. We use the content-based tanh attention described in \cite{vinyals2015grammar}. We use the Mel Encoder described in \cite{hsu2018hierarchical} for Flowtron models that predict the parameters of the Gaussian mixture.

Unlike \cite{ping2017deep,gibiansky2017deep}, where site specific speaker embeddings are used, we use a single speaker embedding that is channel-wise concatenated with the encoder outputs at every token. We use a fixed dummy speaker embedding for models not conditioned on speaker id. Finally, we add a dense layer with a sigmoid output the flow step closest to $\boldsymbol{z}$. This provides the model with a gating mechanism as early as possible during inference to avoid extra computation.

\subsection{Inference}
Once the network is trained, doing inference is simply a matter of randomly sampling $\boldsymbol{z}$ values from a spherical Gaussian, or Gaussian Mixture, and running them through the network, reverting the order of the input when necessary. During training we used $\sigma^2=1$. The parameters of the Gaussian mixture are either fixed or predicted by Flowtron. In section \ref{sec:sampling_prior} we explore the effects of different values for $\sigma^2$.  In general, we found that sampling $\boldsymbol{z}$ values from a Gaussian with a lower standard deviation from that assumed during training resulted in mel-spectrograms that sounded better, as found in ~\cite{kingma2018glow}, and earlier work on likelihood-based generative models~\cite{parmar2018image}. During inference we sampled $\boldsymbol{z}$ values from a Gaussian with $\sigma^2 = 0.5$, unless otherwise specified. The text and speaker embeddings are included at each of the coupling layers as before, but now the affine transforms are inverted in time, and these inverses are also guaranteed by the loss.
\section{Experiments}\label{sec:experiments}
This section describes our training setup and provides quantitative and qualitative results.
Our quantitative results show that Flowtron has mean opinion scores (MOS) that are comparable to that of state of the art models for text to mel-spectrogram synthesis such as Tacotron 2. Our qualitative results display many features that are not possible or not efficient with Tacotron and Tacotron 2 GST. These features include control of the amount of variation in speech, interpolation between samples and style transfer between seen and unseen speakers during training.

We decode all mel-spectrograms into waveforms by using a single pre-trained WaveGlow \cite{prenger2019waveglow} model trained on a single speaker and available on github \cite{mellotron2020github}. During inference we used $\sigma^2=0.7$. In consonance with \cite{valle2019mellotron}, our results suggests that WaveGlow can be used as an universal decoder.

Although we provide images to illustrate our results, they can best be appreciated by listening. Hence, we ask the readers to visit our website~\footnote{\href{https://nv-adlr.github.io/Flowtron}{https://nv-adlr.github.io/Flowtron}} to listen to Flowtron samples.

\subsection{Training setup}
We train our Flowtron, Tacotron 2 and Tacotron 2 GST models using a dataset that combines the LJSpeech (LJS) dataset \cite{ito2017lj} with two proprietary single speaker datasets with 20 and 10 hours each (Sally and Helen). We will refer to this combined dataset as LSH. We also train a Flowtron model on the \textit{train-clean-100} subset of LibriTTS \cite{zen2019libritts} with 123 speakers and 25 minutes on average per speaker. Speakers with less than 5 minutes of data and files that are larger than 10 seconds are filtered out. For each dataset we use at least 180 randomly chosen samples for the validation set and the remainder for the training set. 

The models are trained on uniformly sampled normalized text and ARPAbet encodings obtained from the CMU Pronouncing Dictionary~\cite{weide1998cmu}. We do not perform any data augmentation. We adapt the public Tacotron 2 and Tacotron 2 GST repos to include speaker embeddings as described in Section \ref{sec:audio_flow}.

We use a sampling rate of 22050 Hz and mel-spectrograms with 80 bins using librosa mel filter defaults. We apply the STFT with a FFT size of 1024, window size of 1024 samples and hop size of 256 samples ($\sim 12ms$).

We use the ADAM \cite{kingma2014adam} optimizer with default parameters, 1e-4 learning rate and 1e-6 weight decay for Flowtron and 1e-3 learning rate and 1e-5 weight decay for the other models, following guidelines in \cite{wang2017tacotron}. We anneal the learning rate once the generalization error starts to plateau and stop training once the the generalization error stops significantly decreasing or starts increasing. The Flowtron models with 2 steps of flow were trained on the LSH dataset for approximately 1000 epochs and then fine-tuned on LibriTTS for 500 epochs. Tacotron 2 and Tacotron 2 GST are trained for approximately 500 epochs. Each model is trained on a single NVIDIA DGX-1 with 8 GPUs.

We find it faster to first learn to attend on a Flowtron model with a single step of flow and large amounts of data than multiple steps of flow and less data. After the model has learned to attend, we transfer its parameters to models with more steps of flow and  speakers with less data. Thus, we first train Flowtron model with a single step of flow on the LSH dataset with many hours per speaker. Then we fine tune this model to Flowtron models with more steps of flow. Finally, these models are fine tuned on LibriTTS with an optional new speaker embedding.

\label{sec:quantitative_results}
\subsection{Mean Opinion Score comparison}
We provide results that compare mean opinion scores (MOS) from real data from the LJS dataset, samples from a Flowtron with 2 steps of flow and samples from our implementation of Tacotron 2, both trained on LSH. Although the models evaluated are multi-speaker, we only compute mean opinion scores on LJS. In addition, we use the mean opinion scores provided in \cite{prenger2019waveglow} for ground truth data from the LJS dataset.

We crowd-sourced mean opinion score (MOS) tests on Amazon Mechanical Turk. Raters first had to pass a hearing test to be eligible. Then they listened to an utterance, after which they rated pleasantness on a five-point scale. We used 30 volume normalized utterances from all speakers disjoint from the training set for evaluation, and randomly chose the utterances for each subject.

The mean opinion scores are shown in Table \ref{tab:mos_scores} with $95\%$ confidence intervals computed over approximately 250 scores per source. The results roughly match our subjective qualitative assessment. The larger advantage of Flowtron is in the control over the amount of speech variation and the manipulation of the latent space. 

\begin{table}[!ht]
\begin{center}
    \begin{tabular}{c|c|c|c}
        \textbf{Source} & \textbf{Flows} &  \textbf{Mean Opinion Score (MOS)}\\
        \hline
        Real          & -         & 4.274 $\pm$ 0.1340 \\
        Flowtron      & 3         & 3.665 $\pm$ 0.1634 \\
        Tacotron 2    & -         & 3.521 $\pm$ 0.1721 \\
    \end{tabular}
    \caption{Mean Opinion Score (MOS) evaluations with $95\%$ confidence intervals for various sources.}
    \label{tab:mos_scores}
\end{center}
\end{table}

\subsection{Sampling the prior}\label{sec:sampling_prior}
The simplest approach to generate samples with Flowtron is to sample from a prior distribution $\boldsymbol{z} \sim \mathcal{N}(0, \sigma^2)$ and adjust $\sigma^2$ to control amount of variation. Whereas $\sigma^2=0$ completely removes variation and produces outputs based on the model bias, increasing the value of  $\sigma^2$ will increase the amount of variation in speech. 

\subsubsection{Speech variation}
To showcase the amount of variation and control thereof in Flowtron, we synthesize 10 mel-spectrograms and sample the Gaussian prior with $\sigma^2 \in \{0.0, 0.5, 1.0\}$. All samples are generated conditioned on a fixed speaker \textit{Sally} and text ``\textit{How much variation is there?}" to illustrate the relationship between $\sigma^2$ and variability. 

Our results show that despite all the variability added by increasing $\sigma^2$, all the samples synthesized with Flowtron still produce high quality speech.

\begin{figure}[!ht]
    \centering
    \subfloat[$\sigma^2=0$]{
        \centering
        \includegraphics[trim=100 0 600 0, clip, width=0.85\columnwidth]{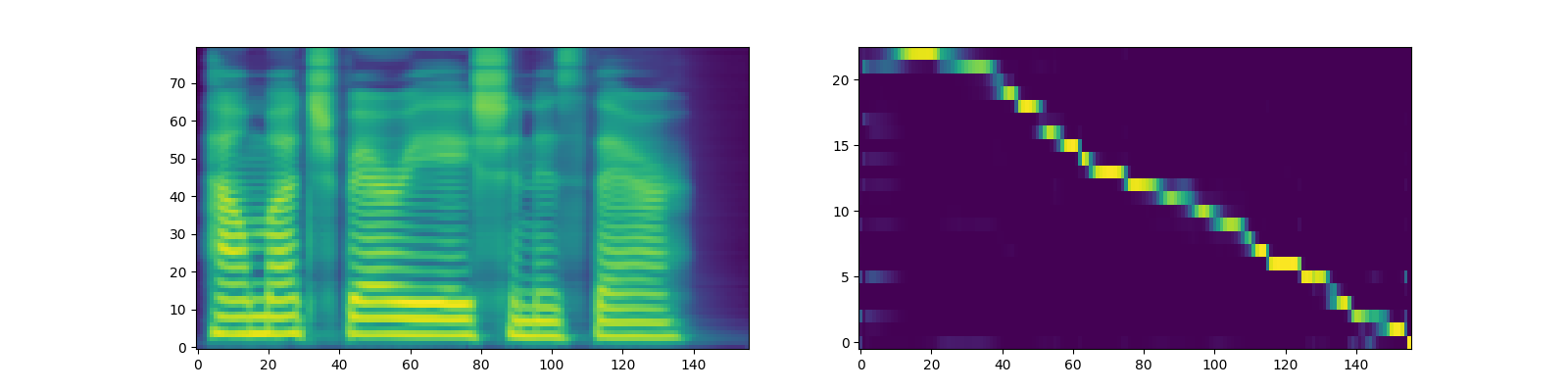}
    }
    
    \subfloat[$\sigma^2=0.5$]{
        \centering
        \includegraphics[trim=100 0 600 0, clip, width=0.85\columnwidth]{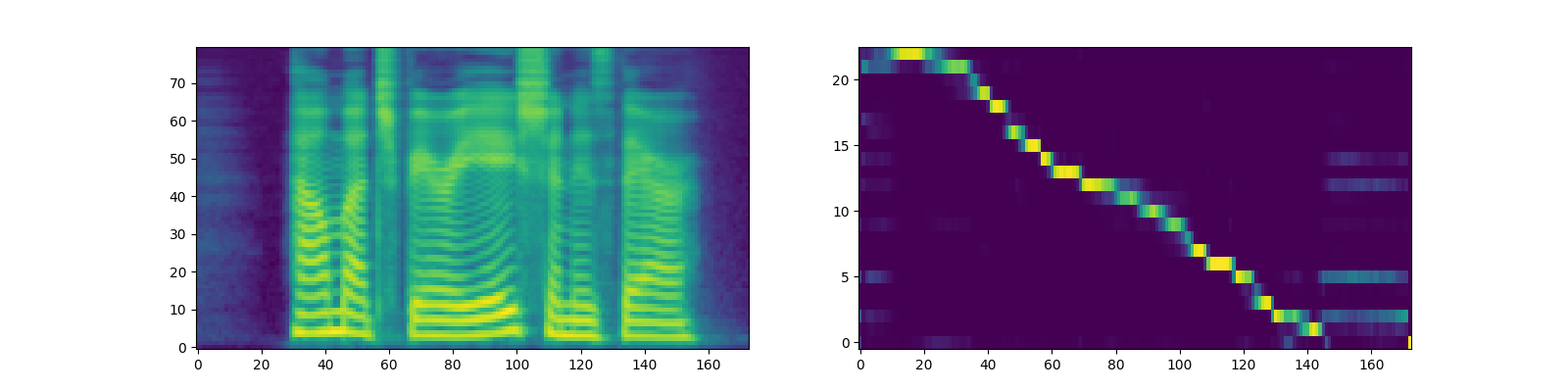}
    } 
    
    \subfloat[$\sigma^2=1$]{
        \centering
        \includegraphics[trim=100 0 600 0, clip, width=0.85\columnwidth]{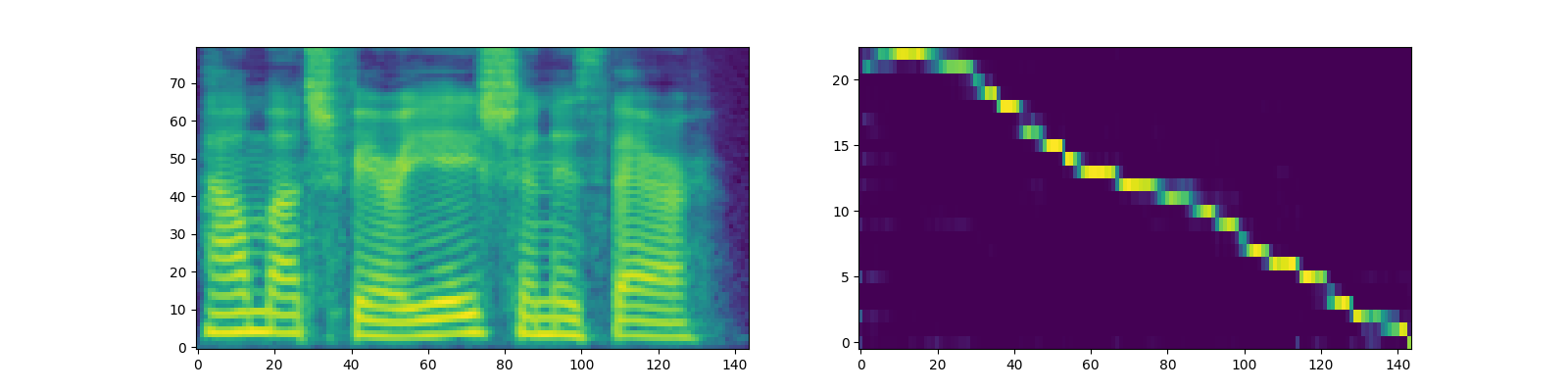}
    }   
    \caption{Mel-spectrograms generated with Flowtron using different $\sigma^2$. This parameter can be adjusted to control mel-spectrogram variability during inference.}
    \label{fig:mel_quality}
\end{figure}

Figure \ref{fig:mel_quality} also shows that unlike most SOTA models \cite{shen2017natural,arik2017deep,arik2017deep2,ping2017deep,skerry2018towards, wang2018style,binkowski2019high}, Flowtron generates sharp harmonics and well resolved formants without a compound loss nor Prenet or Postnet layers.

Now we show that adjusting $\sigma^2$ is a simple and valuable approach that provides more variation and control than Tacotron, without sacrificing speech quality. For this, we synthesize 10 samples with Tacotron 2 using different values for the Prenet dropout probability $p \in \{0.45, 0.5, 0.55\}$. We scale the outputs of the dropout output such that the mean of the output remains equal to the mean with $p=0.5$, the value used during training. Although we also provide samples computed on values of $p \in [0, 1]$ in our supplemental material, we do not include them in our results because they are unintelligible.

\begin{figure}[!ht]
    \centering
    \includegraphics[width=0.9\columnwidth]{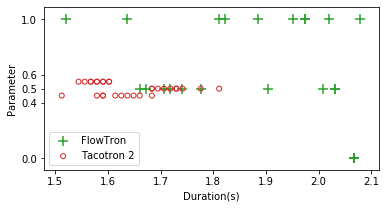}
    \caption{Sample duration in seconds given parameters $\sigma^2$ and $p$. These results show that Flowtron provides more variation in sample duration than Tacotron 2.}
    \label{fig:sentence_durations}
\end{figure}

In Figure \ref{fig:sentence_durations} below we provide scatter plots from sample duration in seconds. Our results show that whereas  $\sigma^2=0$ produces samples with no variation in duration, larger values of $\sigma^2$ produces samples with more variation in duration. Humans manipulate word and sentence length to express themselves, hence this is valuable.

\begin{figure}[!ht]
    \centering
    \subfloat[Flowtron $\sigma^2=0$]{
        \includegraphics[width=\columnwidth]{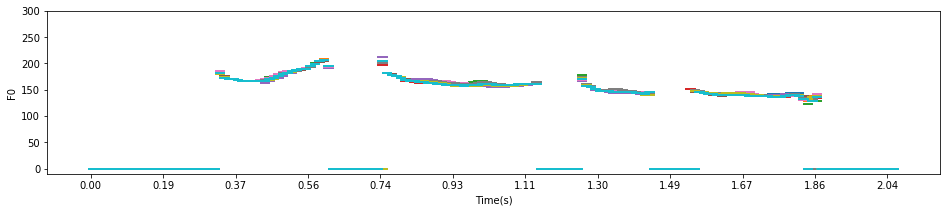}
        \label{fig:inference_sigma0_pitch_contours}
    }
    
    \subfloat[Flowtron $\sigma^2=0.5$]{
        \includegraphics[width=\columnwidth]{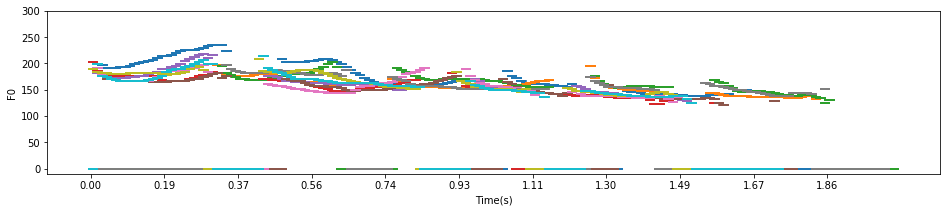}
        \label{fig:inference_sigma0p5_pitch_contours}
    }
    
        \subfloat[Flowtron $\sigma^2=1$]{
        \includegraphics[width=\columnwidth]{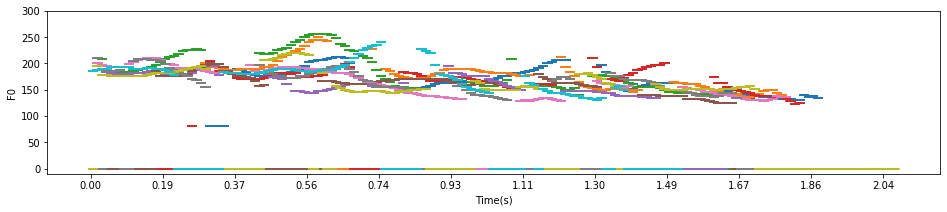}
        \label{fig:inference_sigma1_pitch_contours}
    }
    
     \subfloat[Tacotron 2 $p \in \{0.45, 0.5, 0.55\}$]{
        \includegraphics[width=\columnwidth]{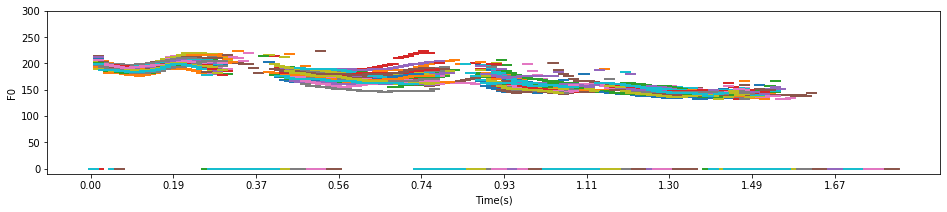}
        \label{fig:inference_tacotron_pitch_contours}
    }
    \caption{$F_0$ contours obtained from samples generated by Flowtron and Tacotron 2 with different values for $\sigma^2$ and $p$. Flowtron provides more expressivity than Tacotron 2.}
    \label{fig:pitch_contours}
\end{figure}

In Figure \ref{fig:pitch_contours} we provide scatter plots of $F_0$ contours extracted with the YIN algorithm \cite{de2002yin}, with minimum $F_0$, maximum $F_0$ and harmonicity threshold equal to 80 Hz, 400 Hz and 0.3 respectively. Our results show a behavior similar to the previous sample duration analysis. As expected, $\sigma^2=0$ provides no variation in $F_0$ contour\footnote{Variations in $\sigma^2=0$ are due to different $\boldsymbol{z}$ for WaveGlow.}, while increasing the value of $\sigma^2$ will increase the amount of variation in $F_0$ contours. 

Our results in Figure \ref{fig:pitch_contours} also show that the samples produced with Flowtron are considerably less monotonous than the samples produced with Tacotron 2. Whereas increasing $\sigma^2$ considerably increases variation in $F_0$, modifying $p$ barely produces any variation. This is valuable because expressive speech is associated with non-monotonic $F_0$ contours.

\subsubsection{Interpolation between samples}
With Flowtron we can perform interpolation in z-space to achieve interpolation in mel-spectrogram space. For this experiment we evaluate Flowtron models with and without speaker embeddings. For the experiment with speaker embeddings we choose the Sally speaker and the phrase ``\textit{It is well known that deep generative models have a rich latent space.}". We generate mel-spectrograms by sampling  $\boldsymbol{z} \sim \mathcal{N}(0, 0.8)$ twice and interpolating between them over 100 steps. 

For the experiment without speaker embeddings we interpolate between Sally and Helen using the phrase ``\textit{We are testing this model.}". First, we perform inference by sampling $\boldsymbol{z} \sim \mathcal{N}(0, 0.5)$ until we find two $\boldsymbol{z}$ values, $\boldsymbol{z_h}$ and $\boldsymbol{z_s}$, that produce mel-spectrograms with Helen's and Sally's voice respectively. We then generate samples by performing inference while linearly interpolating between $\boldsymbol{z_h}$ and $\boldsymbol{z_s}$. 

Our same speaker interpolation samples show that Flowtron is able to interpolate between multiple samples while producing correct alignment maps. In addition, our different speaker interpolation samples show that Flowtron is able to blurry the boundaries between two speakers, creating a speaker that combines the characteristics of both.
\subsection{Sampling the posterior}\label{sec:sampling_posterior}
In this approach we generate samples with Flowtron by sampling a posterior distribution conditioned on prior evidence containing speech characteristics of interest, as described in \cite{gambardella2019transflow,kingma2018glow}. In this experiment, we collect prior evidence $\boldsymbol{z}_{e}$ by performing a forward pass with the speaker id to be used during inference\footnote{To remove this speaker's information from $\boldsymbol{z}_{e}$}, observed mel-spectrogram and text from a set of samples with characteristics of interest. If necessary, we time-concatenate each $\boldsymbol{z}_{e}$ with itself to fulfill minimum length requirements defined according to the text length to be said during inference. 

Tacotron 2 GST \cite{wang2018style} has an equivalent posterior sampling approach, in which during inference the model is conditioned on a weighted sum of global style tokens (posterior) queried through an embedding of existing audio samples (prior). For Tacotron 2 GST, we evaluate two approaches: in one we use a single sample to query a style token in the other we use an average style token computed over multiple samples.

\subsubsection{Seen speaker without alignments}\label{sec:seen_noalignments}
In this experiment we compare Sally samples from Flowtron and Tacotron 2 GST generated by conditioning on the posterior computed over 30 Helen samples with the highest variance in fundamental frequency. The goal is to make a monotonic speaker sound expressive. Our experiments show that by sampling from the posterior or interpolating between the posterior and a standard Gaussian prior, Flowtron is able to make a monotonic speaker gradually sound more expressive. On the other hand, Tacotron 2 GST is barely able to alter characteristics of the monotonic speaker. 

\subsubsection{Seen speaker with alignments}\label{sec:seen_alignments}
We use a Flowtron model with speaker embeddings to illustrate Flowtron's ability to learn and transfer acoustic characteristics that are hard to express algorithmically but easy to perceive acoustically, we select a female speaker from LibriTTS with a distinguished nasal voice and oscillation in $F_0$ as our source speaker and transfer her style to a male speaker, also from LibriTTS, with acoustic characteristics that sound different from the female speaker. Unlike the previous experiment, this time the text and the alignment maps are transferred from the female to the male speaker.

\begin{figure}[!ht]
    \centering
    \subfloat[Female]{
        \centering
        \includegraphics[width=\columnwidth]{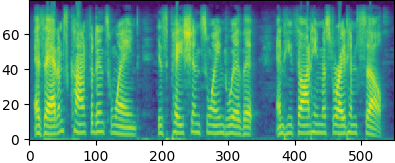}
    }
    
    \subfloat[Transfer]{
        \centering
        \includegraphics[width=\columnwidth]{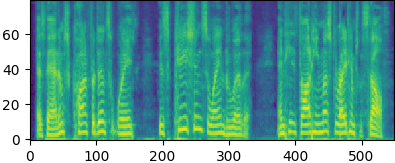}
    }   
    
    \subfloat[Male]{
        \centering
        \includegraphics[width=\columnwidth]{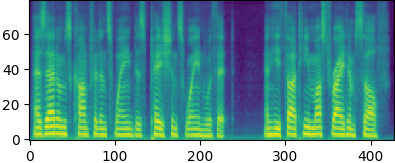}
    }   
    
    \caption{Mel-spectrograms from a female speaker, male speaker and a sample where we transfer the acoustic characteristics from the female speaker to the male speaker. It shows that the transferred sample is more similar to the female speaker than the male speaker.}
    \label{fig:formants}
\end{figure}

Figure \ref{fig:formants} is an attempt to visualize the transfer of these acoustic qualities we described. It shows that after the transfer, the lower partials of the male speaker oscillate more and become more similar to the female speaker.

\subsubsection{Unseen speaker style}\label{sec:unseen_data}
We compare samples generated with Flowtron and Tacotron 2 GST with speaker embeddings in which we modify a speaker's style by using data from the same speaker but from a style not seen during training. Whereas Sally's data used during training consists of news article readings, the evaluation samples contain Sally's interpretation of the somber and vampiresque novel \text{Born of Darkness}.

Our samples show that Tacotron 2 GST fails to emulate the somber style from Born of Darkness's data. We show that Flowtron succeeds in transferring not only to the somber style in the evaluation data, but also the long pauses associated with the narrative style.

\subsubsection{Unseen speaker}\label{sec:unseen_data_and_speaker}
In this experiment we compare Flowtron and Tacotron 2 GST samples in which we transfer the speaking style of a speaker not seen  during training. Both models use speaker embeddings.
 
For these experiments, we consider two speakers. The first comes from speaker ID 03 from RAVDESS, a dataset with emotion labels. We focus on the label ``surprised". The second speaker is Richard Feynman, using a set of 10 audio samples collected from the web. 

For each experiment, we use the Sally speaker and the sentences ``\textit{Humans are walking on the street?}" and ``\textit{Surely you are joking mister Feynman.}", which do not exist in RAVDESS nor in the audio samples from Richard Feynman. 

The samples generated with Tacotron 2 GST are not able to emulate the surprised style from RAVDESS nor Feynman's prosody and acoustic characteristics. Flowtron, on the other hand, is able to make Sally sound surprised, which is drastically different from the monotonous baseline. Likewise, Flowtron is able to pick up on the prosody and articulation details particular to Feynman's speaking style, and transfer them to Sally.

\if
Figure \ref{fig:z_centroids} shows that the Feynman's centroids obtained with each Flowtron model on are considerably different from a sample from a Gaussian distribution, suggesting that the model is able to store speaker specific information, even though the speaker is unseen during training. 

\begin{figure}[!ht]
    \centering
    \subfloat[Flowtron LSH]{
        \centering
        \includegraphics[width=\columnwidth]{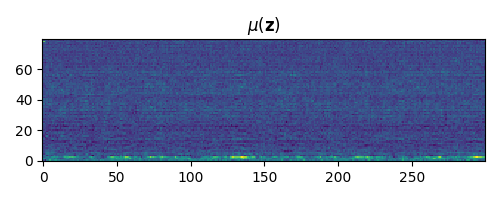}
    }
    
    \subfloat[Flowtron LibriTTS]{
        \centering
        \includegraphics[width=\columnwidth]{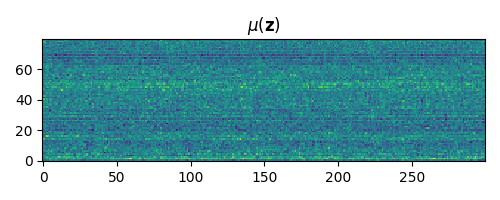}
    }   
    
    \subfloat{
        \centering
        \includegraphics[width=\columnwidth]{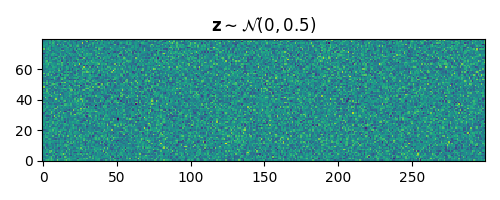}
    }   
    \caption{$\mu(\boldsymbol{z}_{s})$ computed over 10 audio samples using the Flowtron models trained on LSH and LibriTTS. It is clear that the centroids obtained from both Flowtron models have more structure than the centroid obtained from a Gaussian.}
    \label{fig:z_centroids}
\end{figure}
\fi
\subsection{Sampling the Gaussian Mixture}\label{sec:sampling_gm}
In this last section we showcase visualizations and samples from Flowtron Gaussian Mixture (GM). First we investigate how different mixture components and speakers are correlated. Then we provide sound examples in which we modulate speech characteristics by translating one of the the dimensions of an individual component.

\subsubsection{Visualizing assignments}
\begin{figure}[!ht]
    \centering
    \includegraphics[width=\columnwidth]{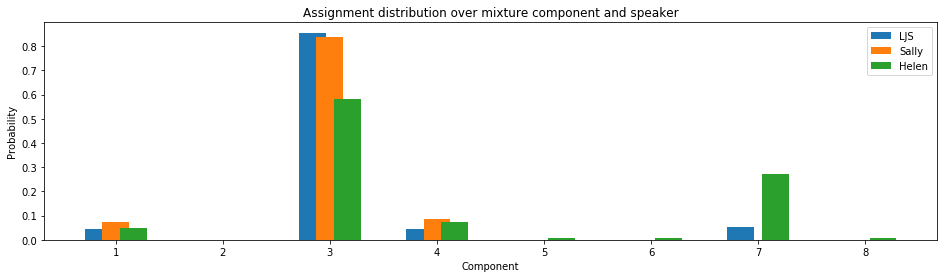}
    \caption{Component assignments for Flowtron GM-A. Unlike LJS and Sally, Helen is almost exclusively assigned to component 7.}
    \label{fig:assignments_a}
\end{figure}

For the first experiment, we train a Flowtrom Gaussian Mixture on LSH with 2 steps of flow, speaker embeddings and fixed mean and covariance (Flowtron GM-A). We obtain mixture component assignments per mel-spectrogram by performing a forward pass and averaging the component assignment over time and samples. Figure \ref{fig:assignments_a} shows that whereas most speakers are equally assigned to all components, component 7 is almost exclusively assigned to Helen's data. 

\begin{figure}
        \centering
        \includegraphics[width=\columnwidth]{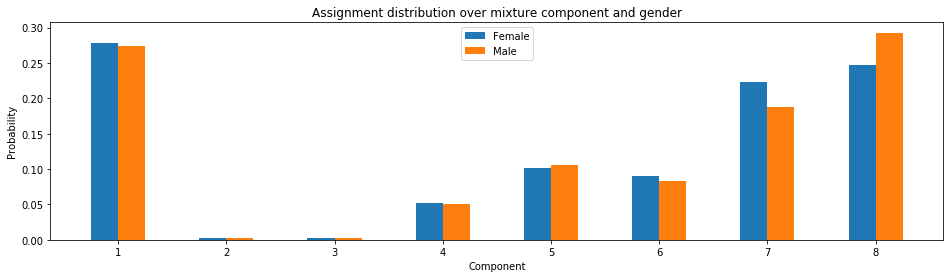}
        \caption{Component assignments for Flowtron GM-B. Components 7 and 8 are assigned different probabilities according to gender, suggesting that the information stored in the components is gender dependent.}
        \label{fig:assignments_b}
\end{figure}

In the second experiment, we train a Flowtron Gaussian Mixture on LibriTTS with 1 step of flow, without speaker embeddings and predicted mean and covariance (Flowtron GM-B). Figure \ref{fig:assignments_b} shows that Flowtron GM assigns more probability to component 7 when the speaker is male than when it's female. Conversely, the model assigns more probability to component 6 when the speaker is female than when it's male.

\subsubsection{Translating dimensions}
In this subsection, we use the model Flowtron GM-A described previously. We focus on selecting a single mixture component and translating one of its dimensions by adding an offset. 

The samples in our supplementary material show that we are able to modulate specific speech characteristics like pitch and word duration. Although the samples generated by translating one the dimensions associated with pitch height have different pitch contours, they have the same duration.  Similarly, our samples show that translating the dimension associated with length of the first word does not modulate the pitch of the first word. This provides evidence that we can modulate these attributes by manipulating these dimensions and that the model is able to learn a disentangled representation of these speech attributes.

\section{Discussion}\label{sec:discussion}
In this paper we propose a new text to mel-spectrogram synthesis model based on autoregressive flows that is optimized by maximizing the likelihood and allows for control of speech variation and style transfer. Our results show that samples generated with FlowTron achieve mean opinion scores that are similar to samples generated with state-of-the-art text-to-speech synthesis models. In addition, we demonstrate that at no extra cost and without a compound loss term, our model learns a latent space that stores non-textual information. Our experiments show that FlowTron gives the user the possibility to transfer charactersitics from a source sample or speaker to a target speaker, for example making a monotonic speaker sound more expressive. 

Our results show that despite all the variability added by increasing $\sigma^2$, the samples synthesized with FlowTron still produce high quality speech. Our results show that FlowTron learns a latent space over non-textual features that can be investigated and manipulated to give the user more control over the generative model’s output. We provide many examples that showcase this including increasing variation in mel-spectrograms in a controllable manner, transferring the style from speakers seen and unseen during training to another speaker using sentences with similar or different text, and making a monotonic speaker sound more expressive. 

Flowtron produces expressive speech without labeled data or ever seeing expressive data. It pushes text-to-speech synthesis beyond the expressive limits of personal assistants. It opens new avenues for speech synthesis in human-computer interaction and the arts, where realism and expressivity are of utmost importance. To our knowledge, this work is the first to demonstrate the advantages of using normalizing flow models in text to mel-spectrogram synthesis.

\newpage
\bibliography{main}
\bibliographystyle{icml2020}

\newpage
\end{document}